**Coupling Spin Defects in a Layered Material to Nanoscale Plasmonic Cavities**

Noah Mendelson[1], Ritika Ritika[1], Mehran Kianinia[1,2], John Scott[1,2], Sejeong Kim[3], Johannes E. Fröch[1], Camilla Gazzana[1], Mika Westerhausen[1], Licheng Xiao[4, 5], Seyed Sepehr Mohajerani[4, 5], Stefan Strauf[4, 5], Milos Toth[1,2], Igor Aharonovich[1,2]* and Zai-Quan Xu[1]*

[1]School of Mathematical and Physical Sciences, University of Technology Sydney, Ultimo, New South Wales 2007, Australia
[2]ARC Centre of Excellence for Transformative Meta-Optical Systems (TMOS), University of Technology Sydney, Ultimo, New South Wales 2007, Australia
[3]Department of Electrical and Electronic Engineering, University of Melbourne, Victoria 3010, Australia
[4]Department of Physics, Stevens Institute of Technology, Hoboken, New Jersey 07030, USA
[5]Center for Quantum Science and Engineering, Stevens Institute of Technology, Hoboken, New Jersey 07030, USA

*Corresponding Authors
E-mail: igor.aharonovich@uts.edu.au and zaiquan.xu@uts.edu.au.

**Abstract**
Spin defects in hexagonal boron nitride, and specifically the negatively charged boron vacancy ($V_B^-$) centres, are emerging candidates for quantum sensing. However, the $V_B^-$ defects suffer from low quantum efficiency and as a result exhibit weak photoluminescence. In this work, we demonstrate a scalable approach to dramatically enhance the $V_B^-$ emission by coupling to a plasmonic gap cavity. The plasmonic cavity is composed of a flat gold surface and a silver cube, with few-layer hBN flakes positioned in between. Employing these plasmonic cavities, we extracted two orders of magnitude in photoluminescence enhancement associated with a corresponding 2-fold enhancement in optically detected magnetic resonance contrast. The work will be pivotal to progress in quantum sensing employing 2D materials, and realisation of nanophotonic devices with spin defects in hexagonal boron nitride.

Hexagonal boron nitride (hBN) has recently emerged as an attractive layered material for nanophotonics[1]. Some of its unique features include strong confinement of phonon polaritons[2] and a wide bandgap[3] that facilitates a range of deep fluorescent defects that act as single photon emitters[4-6]. Combined, these properties offer unique opportunities to explore nanoscale light matter interaction phenomena and engineer on chip nanophotonic devices.

From the growing family of defects in hBN, the negatively charged boron vacancy ($V_B^-$) is of particular interest because of its recently-discovered spin properties[7-14]. The spin-photon interface enables emerging applications in quantum information, and more specifically nanoscale quantum sensing and quantum imaging[15, 16]. In particular, the $V_B^-$ defects has a triplet ground state with ground state splitting of ~ 3.5 GHz and its spin state can be optically readout and coherently manipulated. A schematic of a $V_B^-$ defect in hBN is shown in Figure 1a and its level structure is shown in Figure 1b. These properties constitute its potential as an emerging quantum spin sensor for strain, temperature and ambient magnetic fields in van der Waals crystals[9]. Furthermore, these defects can be generated deterministically by neutron irradiation, focused ion beam (FIB) irradiation[17] or laser ablation[10, 11], which is important for their future employment in scalable devices. The basic optical properties of the $V_B^-$ are shown in figure 1c. It has a broad emission at around ~ 850 nm, that is slightly reduced at 4K[17]. The extracted lifetimes, as shown in Figure 1d, are 1.73 and 3.80 ns at room temperature (red) and 4K (blue), respectively. This suggests a reduction of non-radiative recombination at cryogenic temperature[18, 19].

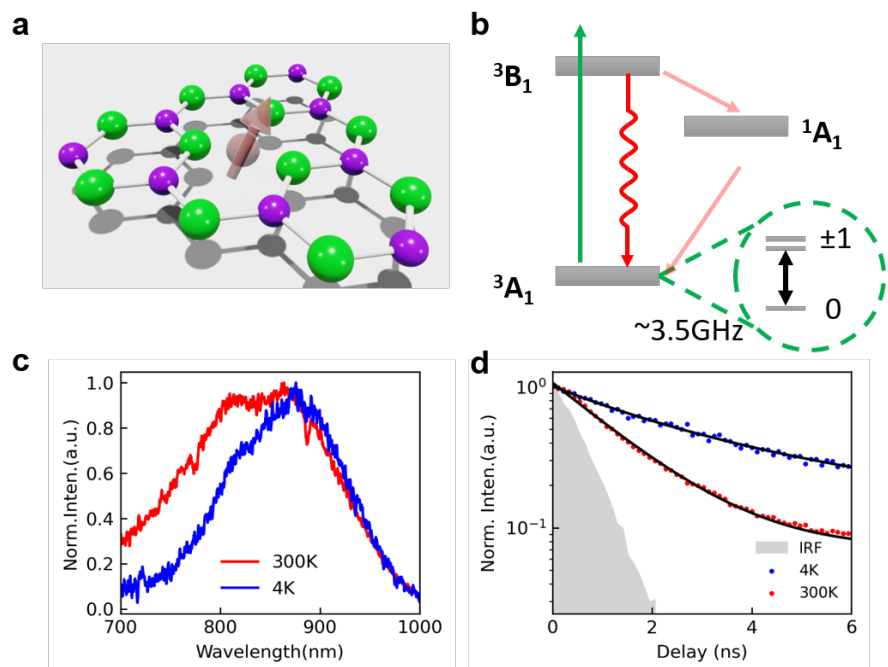

Figure 1. (a) 3D illustration of the boron vacancy ($V_B^-$) in hBN, with boron and nitrogen atoms presented as purple and green spheres, respectively. The vacancy resembles a single electron system in a solid material associated with a spin, which can be read out optically. (b) Level structure of the $V_B^-$. (c) Normalized PL spectra of hBN $V_B^-$ vacancy at 300K (red) and 4 K (blue). (d) Time-dependent PL spectra of $V_B^-$ at 300K (red) and 4K (blue). The extracted lifetime values

*are 1.73 ns and 3.80 ns, respectively. The instrument response function (IRF) is shaded in grey, yielding an IRF of 0.77 ns.*

A major limitation of the $V_B^-$ defects is their low brightness that originates from a high intrinsic non-radiative decay rate and results in a rather low optically detected magnetic resonance (ODMR) contrast[7, 13]. To circumvent the issue of low quantum yield, emitters can be coupled to plasmonic cavities hat enhance the excitation fields as well as the emission rates. Such an approach have been employed to enhance emission from variety of systems, including transition metal di-chalcogenides and defects in solid[20-24]. Given the broad emission of the $V_B^-$ defects, plasmonic gap cavities offer an excellent solution for emission enhancement[25]. In addition, these cavities require very thin (below ~ 10 nm) medium to host the light source, which are particularly appealing for the case of a layered material such as hBN. Finally, the plasmonic gap cavities also feature a broad, tunable resonance, extremely small mode volumes, easy nanofabrication methods and can operate at room-temperature[25-29].

In the current work, we demonstrate coupling of an ensemble of $V_B^-$ emitters to a plasmonic gap caviti. Specifically, we employ a gold mirror and a silver nanocube to construct the plasmonic gap cavity. To study the effect of the assembly, we investigate two complementary strategies: (1) hBN is transferred onto a gold mirror and the silver cubes are positioned on top of the hBN layer, and (2) hBN is transferred onto the silver cubes and capped with a synthetic ultra-flat gold mirror. Both methods yielded significant enhancement of photoluminescence (PL) emission from the $V_B^-$ defect, accompanied by reduction in fluorescent lifetimes and improvement in the ODMR contrast. Our findings constitute an important basis for further studies of plasmon coupling to defects in 2D materials and will benefit future development of quantum sensing using this quantum spin system.

The general approach of the plasmonic gap cavities integrated with $V_B^-$ defects is shown schematically in figure 2a. This plasmonic structure confines light to the gap between the nanocubes and the gold mirror, which ensures optimal spatial matching of the defects with the hot spot of the gap cavity. This type of cavity significantly improves emission intensity by the combined effect of increased excitation efficiency, enhanced emission rate and redistribution of the radiation pattern into a narrow range of the far field which enables an improved collection efficiency. We used a nanoplasmonic gap cavity cosnsisting of an ultraflat gold mirror (~ 90 nm, covered by 2 nm $Al_2O_3$), hBN flakes with ~ 8 nm thickness and a silver/gold nanocube (~ 100 nm edge length). To generate optically active $V_B^-$ defects, hBN flakes were irradiated homogeneously using a nitrogen FIB with an optimized dose (see Figure S1 and the Methods section). The resonance of this plasmonic gap cavity is broad, as indicated by the scattering spectra shown in Figure 2b. The resonance spans a spectral range from 600 to 850 nm with an emission maximum at ~ 700 nm.

The gap cavity system was analysed further using finite-difference time-domain methods. Both hBN layer (8 nm) and $Al_2O_3$ (2 nm) layers are included and the dipole emitter with vertical linear polarization is located in the middle of the hBN layer. Figure 2c shows the side-view of the electric field profile of the mode. The calculated PL enhancement and Purcell factors as a function of gap thickness are plotted in Figure 2d, which was varied from 4 to 12 nm. Typically, the Purcell factor decreases with increasing gap size due to an increase in mode volume. However, in this case, the resonance centred on 800 nm has a maximum at a gap thickness of ~ 8 nm, resulting in the highest Purcell factor and maximum PL enhancement.

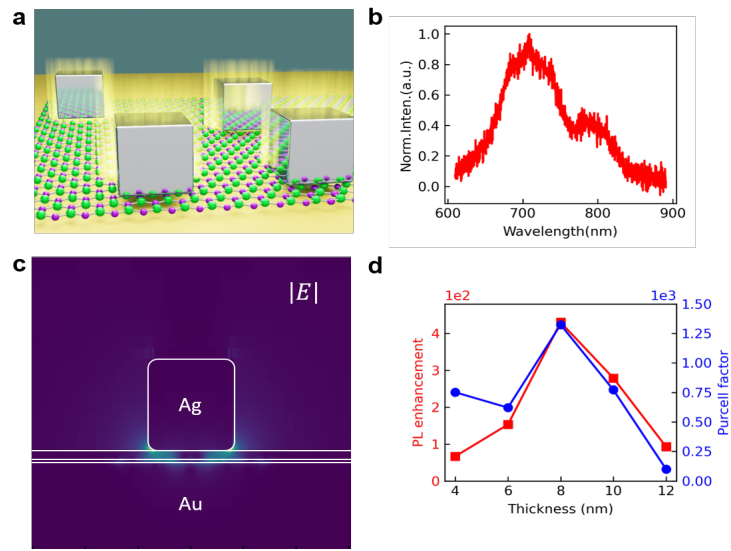

*Figure 2. Integration of $V_B^-$ emitters in a plasmonic gap cavity. (a) Schematic illustration of a nanocube-on-mirror plasmonic gap cavity. FIB irradiated hBN flakes are placed inside the gap to spatially match the cavity hot spot. (b) Experimental optical scattering spectrum of a single nanocavity. (c) Electric field profile for gap cavity at λ = 800 nm with 2 nm $Al_2O_3$ and 8 nm hBN. (d) calculated PL enhancement and Purcell factor versus hBN thickness.*

Comprehensive studies were performed to compare the emission from uncoupled (as-fabricated) $V_B^-$ ensembles to those integrated in gap cavities using a home-built confocal optical microscope. A side-view of the sample configuration is presented in Figure 3a, and Figure 3b shows a PL intensity map of the fabricated sample, in which the red spots suggest enhanced emission due to coupling to gap cavities. The white circle outlines the spot we used to illustrate the characteristics of this system. Figure 3c shows the PL spectrum from this spot (coupled: blue) compared with a spectrum from the uncoupled $V_B^-$ emission (red) taken from the same flake. The spectral peak height is ~ 17 times greater when the $V_B^-$ ensemble is coupled to the plasmonic gap cavity. SEM and the dark-field image of the sample (Figure S2) demonstrate there is only one silver nanocube under the laser spot. Plots of intensity measured as a function of excitation power are shown for both cases in Figure 3d. The intensity was integrated from 715 to 900 nm, and the results were fitted using the following equation

$$I = I_{sat}P/(P + P_{sat}),$$

where the fit parameters $I_{sat}$ and $P_{sat}$ are saturation intensity and power, respectively, which are 0.184 and 2.497 MHz, and 7.738 and 8.825 mW for the uncoupled and coupled cases, respectively. Throughout the investigated power range, the PL enhancement factor is ~ 15, on par with the measured example shown above. This is underlined further by the fact that the PL intensity enhancement is approximately constant throughout the power range, and a minor change in the saturation power caused by coupling to a gap cavity.

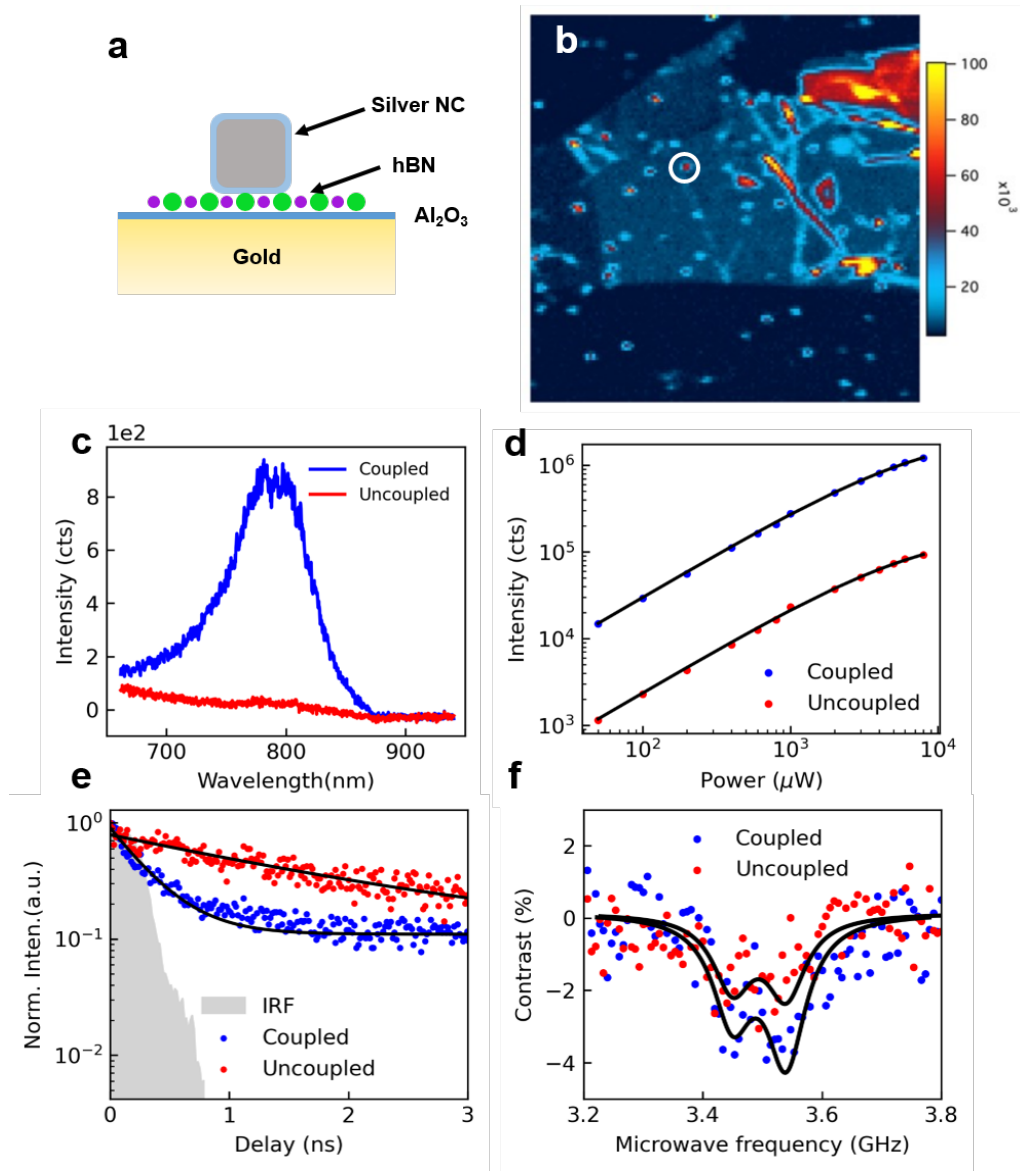

*Figure 3. (a) A side-view of the conventional gap plasmonic sample structure used. (b) Confocal PL intensity (integrating from 715 to 900 nm )mapping of the area of interest. The white circle highlights a spot where $V_B^-$ is coupled in the plasmonic resonance. (c) PL spectra of $V_B^-$ coupled (blue) and uncoupled (red) to the gap cavity. The sample is excited with a 532nm CW laser. (d) Integrated PL intensity of the VB as a function of excitation power. (e) Time-dependent PL spectra of $V_B^-$ coupled (blue) and uncoupled (red) to the gap cavity. The extracted lifetime values are 0.31 and 1.80 ns, respectively. The IRF is shaded in grey. (f) ODMR spectrum of the $V_B^-$ coupled (blue) and uncoupled (red) to the gap cavity.*

To delineate the contributing enhancement factors, we further performed PL decay rate measurements, as is shown in Figure 3e. The lifetimes are determined to be 1.80 ns for $V_B^-$ on $SiO_2$ substrates, that is reduced further to 0.31 ns, for a cavity-integrated ensemble suggesting the decay rate is increased by a factor of ~ 6 when the $V_B^-$ emission is in resonance with the cavity. That indicates that the emission enhancement we observe experimentally consists of both excitation enhancement as well as emission enhancement via the Purcell effect, as will be discussed below.

Finally, we measured the ODMR spectra to analyse the spin readout properties of the coupled system. A microwave field was applied through a copper wire (~20 µm in diameter), placed within 10 µm of the hBN flake. The PL counts were recorded when the microwave was swept from 3.1 to 3.8 GHz. 25 ODMR spectra were integrated and the results are shown in Figure 3f. Two dips at 3.45 and 3.55 GHz are resolved for both the uncoupled and coupled $V_B^-$ emissions, respectively, with fits of the spectra obtained using two Lorentzian functions. Notably, the PL emission enhancement improves the signal-to-background ratio of ODMR spectra, i.e. the ODMR signal from $V_B^-$ in the cavity displays an optical contrast of 4.5% as compared to 2.5% for the pristine ensemble. More statistical data of the enhancement is shown in figure S4.

Next, we explore an alternative version of the plasmonic gap coupled system with an inverted geometry. The inverted structure preferentially will induce strain at the edges of the silver cubes, were the electric-fields are the most intense. In these configuration, strain activated emitters[30] can be formed from hBN and other TMDCs[31] and hence this geometry should be explored.

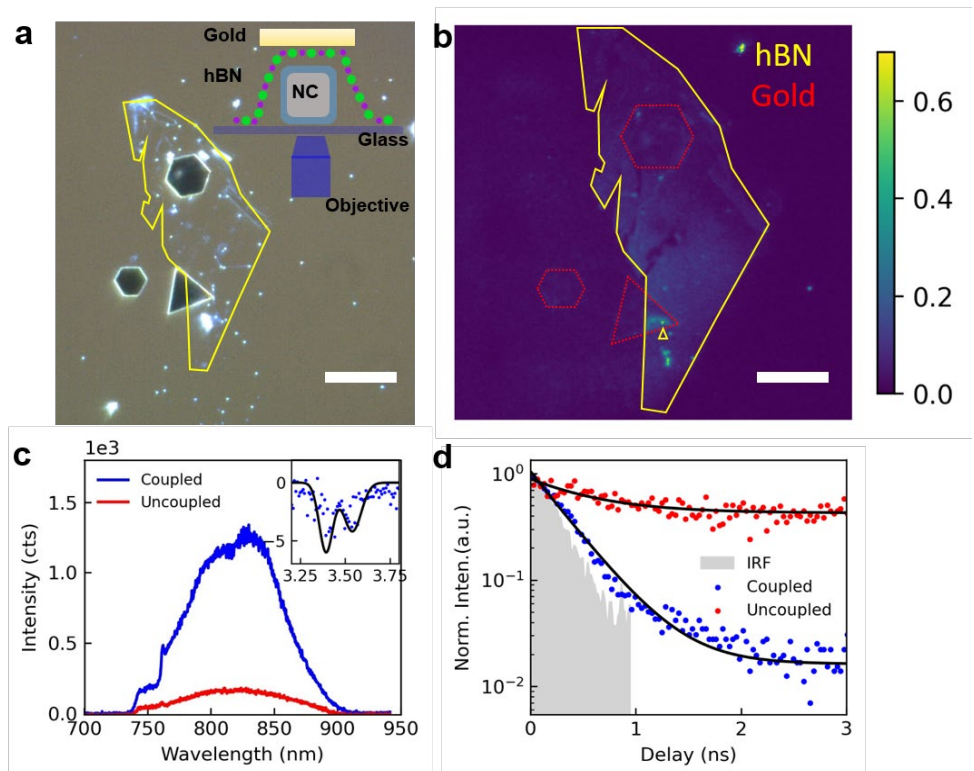

*Figure 4. (a) Dark field optical image of the fabricate hBN in gap plasmonic structures. The hBN flake is transferred onto a glass cover slip and outlined in yellow. The black triangle and hexagons are the ultra-flat gold crystals. The white dots are the silver nanocubes. Inset: a side-view of the inverted sample structure used. (b) Confocal PL intensity mapping of the corresponding area. The yellow triangle points to a spot where $V_B^-$ is coupled in the plasmonic*

*resonance. Scale bar: 15 μm. (c) PL spectra of $V_B^-$ on and off gap cavity. Inset: ODMR spectrum of the $V_B^-$ resonates with the plasmonic cavity. (d) Time-dependent PL spectra of VB on and off gap cavity. The extracted lifetime values are 0.56 and 1.30 ns, respectively. The instrument response function (IRF) is shaded in grey, yielding an IRF of 0.27 ns.*

To engineer these devices, silver nanocubes were first drop-cast on a glass coverslip with a thickness of 180 μm. Thin hBN flakes, were transferred onto the nanocubes and irradiated with nitrogen to engineer the $V_B^-$ ensembles. Then, the synthesised gold crystals were transferred on top of the hBN with a deterministic align transfer method to complete the cavity. A schematic illustration of the sample structure is shown as an inset in Figure 4a. An inverted scanning optical microscope with an oil-immersion objective (NA = 1.3) was used to characterise the $V_B^-$ emitters in this configuration. A dark-field optical microscope image of the sample is shown in Figure 4a. Bright spots correspond to the locations of the silver nanocubes, and the dark triangle and hexagons are the gold crystals. A PL intensity map is shown in Figure 4b, in which an enhanced spot in the plasmonic gap cavity (coupled $V_B^-$) is marked with a small yellow triangle.

PL spectra recorded from the uncoupled (red) and coupled (blue) $V_B^-$ ensembles are shown in Figure 4c. The peak height is enhanced by ~7 times while the peak position remains nearly identical. The ODMR spectrum shown as an insert in Figure 4c shows an optical contrast of ~ 5%, which is comparable to the cavity-coupled $V_B^-$ in the geometry shown in Figure 3. This level of contrast therefore also suggests that the enhancement originated from $V_B^-$ coupled to the plasmonic gap cavity. We record a lifetime of 0.56 and 1.30 ns for cavity-coupled and uncoupled $V_B^-$, respectively, illustrating an enhancement factor of 2 in emission rate (Figure 4d).

To quantify the PL enhancement, we define the average PL enhancement factor across a single cavity as: <EF> = ($I_{coupled}/I_{uncoupled}$) x ($A_0/A_{cav}$) where $I_{coupled}$ is the PL intensity from the $V_B^-$ coupled into the cavity and $I_{uncoupled}$ is the PL intensity from $V_B^-$ on thermal oxide, and $A_0$ and $A_{cav}$ are the areas of the laser spot and the cavity, respectively. Here we assume that the laser spot is diffraction limited ~ $(266 \text{ nm})^2$ and the cavity size is $(\sim 100 \text{ nm})^2$, yielding a maximum <EF> of 106.1 for the conventional (Figure 3) 70.7 for the inverted (Figure 4) structure, respectively, at an excitation wavelength of 532 nm. Consequently, we can derive the Purcell effect which is defined by the ratio of the radiative rate into the cavity mode to the free-space radiative rate. In the case of the $V_B^-$ system, the non-radiative decay pathways dominate, thus one would expect only a marginal change in lifetime[32]. Hence, we consider the quantum yield to be less than 1% and determine the Fp using the following equation[32, 33], QY x $F_p$ = ( $\tau_{uncoupled}$ / $\tau_{coupled}$)-1 where $\tau_{uncoupled}$ and $\tau_{coupled}$ are the extracted lifetime and QY is the quantum yield. These give us an Fp value of ~ 480 and ~ 130, respectively, for the conventional and inverted geometries.

Despite the significant emission enhancement and the overall high Purcell values, we did not manage to isolate a single $V_B^-$. Using the inverted geometry, we expected the localisation of small ensemble of $V_B^-$ to be concentrated at the edge of the nanocube, and only a selected few to be enhanced. However, second order auto-correlation function measurements did not reveal any non-classical emission from the ensemble.

To conclude, we have demonstrated the integration of the defects in hBN with plasmonic nanocavities. We studied two complementary geometries of silver cubes on top and below the hBN, enclosed by a gold mirror. Both geometries exhibit a major enhancement in PL emission, associated with a modest lifetime reduction and an improved ODMR contrast. Improvement in engineering of the $V_B^-$ and a better nanoscale position should eventually result in isolation of single $V_B^-$ defects in hBN. Overall, our study paves the way to exciting developments in quantum sensing applications employing van der Waals materials.

During the manuscript preparation we became aware of a complementary work reporting on plasmonic enhancement from the $V_B$ defects in hBN[34].

**Acknowledgments**

The authors acknowledge financial support from the Australian Research Council (CE200100010) and the Asian Office of Aerospace Research & Development (FA2386-20-1-4014). The authors thank the Australian Nanofabrication Facilities at the UTS OptoFab node.